\documentclass[%
 twocolumn,
 superscriptaddress,
 amsmath,amssymb,
 aps,
 pra,
 floatfix,
]{revtex4-2}

\usepackage{graphicx}
\usepackage{dcolumn}
\usepackage{bm}
\usepackage[colorlinks,linkcolor=blue,citecolor=blue,urlcolor=blue]{hyperref}

\usepackage{cases}

\usepackage{color}
\definecolor{bluereply}{RGB}{61,112,200}
\definecolor{redreply}{RGB}{220,0,0}
\definecolor{greenreply}{RGB}{0,200,0}

\begin{document}

\preprint{APS/123-QED}

\title{High-dimensional multi-input quantum random access codes and mutually unbiased bases}

\author{Rui-Heng Miao}
\thanks{R.-H. Miao and Z.-D. Liu contributed equally to this work}
\affiliation{CAS Key Laboratory of Quantum Information, University of Science and Technology of China, Hefei 230026, China}
\affiliation{CAS Center For Excellence in Quantum Information and Quantum Physics, University of Science and Technology of China, Hefei 230026, China}
\affiliation{Hefei National Laboratory, University of Science and Technology of China, Hefei 230088, China}

\author{Zhao-Di Liu}
\email{zdliu@ustc.edu.cn}
\affiliation{CAS Key Laboratory of Quantum Information, University of Science and Technology of China, Hefei 230026, China}
\affiliation{CAS Center For Excellence in Quantum Information and Quantum Physics, University of Science and Technology of China, Hefei 230026, China}
\affiliation{Hefei National Laboratory, University of Science and Technology of China, Hefei 230088, China}

\author{Yong-Nan Sun}
\email{synan@hznu.edu.cn}
\affiliation{Department of Physics, Hangzhou Normal University, 311121 Hangzhou, Zhejiang, China}
\affiliation{CAS Key Laboratory of Quantum Information, University of Science and Technology of China, Hefei 230026, China}
\affiliation{CAS Center For Excellence in Quantum Information and Quantum Physics, University of Science and Technology of China, Hefei 230026, China}

\author{Chen-Xi Ning}
\affiliation{CAS Key Laboratory of Quantum Information, University of Science and Technology of China, Hefei 230026, China}
\affiliation{CAS Center For Excellence in Quantum Information and Quantum Physics, University of Science and Technology of China, Hefei 230026, China}
\affiliation{Hefei National Laboratory, University of Science and Technology of China, Hefei 230088, China}

\author{Chuan-Feng Li}
\email{cfli@ustc.edu.cn}
\affiliation{CAS Key Laboratory of Quantum Information, University of Science and Technology of China, Hefei 230026, China}
\affiliation{CAS Center For Excellence in Quantum Information and Quantum Physics, University of Science and Technology of China, Hefei 230026, China}
\affiliation{Hefei National Laboratory, University of Science and Technology of China, Hefei 230088, China}

\author{Guang-Can Guo}
\affiliation{CAS Key Laboratory of Quantum Information, University of Science and Technology of China, Hefei 230026, China}
\affiliation{CAS Center For Excellence in Quantum Information and Quantum Physics, University of Science and Technology of China, Hefei 230026, China}
\affiliation{Hefei National Laboratory, University of Science and Technology of China, Hefei 230088, China}

\date{\today}

\begin{abstract}
Quantum random access codes (QRACs) provide a basic tool for demonstrating the advantages of quantum resources and protocols, which have a wide range of applications in quantum information processing tasks. However, the investigation and application of high-dimensional $(d)$ multi-input $(n)$ $n^{(d)}\rightarrow1$ QRACs are still lacking. Here, we present a general method to find the maximum success probability of $n^{(d)}\rightarrow1$ QRACs. In particular, we give the analytical solution for maximum success probability of $3^{(d)}\rightarrow1$ QRACs when measurement bases are mutually unbiased bases (MUBs). Based on the analytical solution, we show the relationship between MUBs and $n^{(d)}\rightarrow1$ QRACs. First, we provide a systematic method of searching for the operational inequivalence of MUBs (OI-MUBs) when the dimension $d$ is a prime power. Second, we theoretically prove that, surprisingly, the commonly used Galois MUBs are not the optimal measurement bases to obtain the maximum success probability of $n^{(d)}\rightarrow1$ QRACs, which indicates a breakthrough according to the traditional conjecture regarding the optimal measurement bases. Furthermore, based on high-fidelity high-dimensional quantum states of orbital angular momentum, we experimentally achieve two-input and three-input QRACs up to dimension 11. We experimentally confirm the OI-MUBs when $d=5$. Our results open alternative avenues for investigating the foundational properties of quantum mechanics and quantum network coding.
\end{abstract}

\maketitle


\section{Introduction}
Quantum resources are known to outperform their classical counterparts in a large variety of communication tasks. For instance, quantum superdense coding can be used to transfer two classical bits of information by transmitting a single two-level quantum system with the aid of entanglement \cite{Bennett1992}. However, in the absence of entanglement, quantum resources and protocols might not be better than their classical counterparts. For example, the well-known Holevo bound \cite{Holevo1973} places a restriction on the amount of classical information that can be extracted from a quantum state, which might imply that quantum information is no more powerful than classical information. In actuality, quantum random access codes (QRACs) play a key role in quantum information theory to demonstrate whether quantum information is more powerful than classical information. QRACs are communication protocols that enable the compression of an $n$-dit string into one qudit, such that one can recover one of the $n$ dits with high probability ($n^{(d)}\rightarrow1$ QRACs) \cite{Bell1964,Paw2009}. QRACs were originally introduced in the context of quantum finite automata \cite{Ambainis1999, Ambainis2002}. Subsequently, QRACs have been adapted to quantum communication complexity \cite{Klauck2001, Aaronson2004}, especially for locally decodable codes \cite{Kerenidis2004} and network coding \cite{Hayashi}. QRACs have also been applied for semi-device-independent (SDI) quantum randomness extraction \cite{Li2011, Lunghi2015, Li2015} as well as SDI key distribution \cite{Pawlowski2011}.

Recently, high-dimensional QRACs have attracted considerable attention. For example, Tavakoli $et$ $al.$ have given a strategy of $2^{(d)}\rightarrow1$ random access codes and QRACs \cite{Tavakoli2015}, which is proved to be the optimal strategy in \cite{Andris2015,Farkas2019}, showing that high-level quantum systems provide significant advantages over their classical counterparts. In particular, QRACs can provide SDI tests for the detection of measurement incompatibility \cite{Anwer2020, Carmeli2020}, construct SDI self-tests for mutually unbiased bases (MUBs) in arbitrary dimensions \cite{Farkas2019}, and expand to entanglement-assisted QRACs \cite{Wang2019}.

As shown in previous works \cite{Tavakoli2015,Anwer2020, Carmeli2020,Farkas2019,Aguilar2018,Wang2019}, MUBs play a central role in QRACs. For instance, the maximum success probability of $2^{(d)}\rightarrow1$ QRACs is obtained only if the measurement bases are mutually unbiased; this has been strictly proven in reference \cite{Farkas2019}. Furthermore, $2^{(d)}\rightarrow1$ QRACs have been used to investigate whether a special number of MUBs exist in a given dimension \cite{Aguilar2018}. However, determining the relationship between $n^{(d)}\rightarrow1$ QRACs and MUBs is a problem that remains to be addressed. A natural question is whether the maximum success probability of $n^{(d)}\rightarrow1$ QRACs ($n \geqslant 3$) can be achieved through MUBs, and the answer to this question remains unknown, even though some work has conjectured that MUBs are the optimal strategy \cite{Tavakoli2015,Aguilar2018}. The choice of subsets of MUBs will affect the results of quantum information tasks; Hiesmayr \emph{et al.} found detecting entanglement can be more effective with inequivalent MUBs \cite{Hiesmayr2021}. $n^{(d)}\rightarrow1$ QRACs can also show the differences, called operational inequivalence of MUBs (OI-MUBs), which proves at least not all MUBs can achieve the maximum success probability \cite{Aguilar2018}. But to date, no systematic method has been presented that could help predict the OI-MUBs.

In this paper, we provide a general solution for obtaining the maximum success probability of $n^{(d)}\rightarrow1$ QRACs. To give a more in-depth description of their properties and applications, we focus on the case of $3^{(d)}\rightarrow1$ QRACs when measurement bases are MUBs and present the analytical solution using two ways. We carefully investigate the relationship between MUBs and $n^{(d)}\rightarrow1$ QRACs. Following the analytical solution, we introduce a pattern through which it is possible to conjecture when OI-MUBs is present with the dimension $d$ being a prime power. We provide numerical proof of such operational inequivalence up to dimension 1000 (100) when $d$ is a prime (prime power). We demonstrate that, surprisingly, the commonly used Galois MUBs are not the optimal measurement bases for $n^{(d)}\rightarrow1$ QRACs. Here Galois MUBs represent the complete $d+1$ MUBs constructed by a Galois field \cite{Wootters1989,Durt2010}. This may open an alternative direction for researching the optimal bases. Finally, we experimentally demonstrate the advantage of $n^{(d)}\rightarrow1$ QRACs based on the high-dimensional orbital angular momentum (OAM) states of photons. In particular, we experimentally realize $2^{(d)}\rightarrow1$ and $3^{(d)}\rightarrow1$ QRACs up to dimension $11$; we also confirm the OI-MUBs when $d=5$.

\section{\texorpdfstring{Maximum success probability of $\MakeLowercase{n}^{(\MakeLowercase{d})}{\longrightarrow}1$ QRACs}{Maximum success probability of n\^{}(d)->1 QRACs}}
\begin{figure}[tbph]
	\centering
	\includegraphics[width=3.3in]{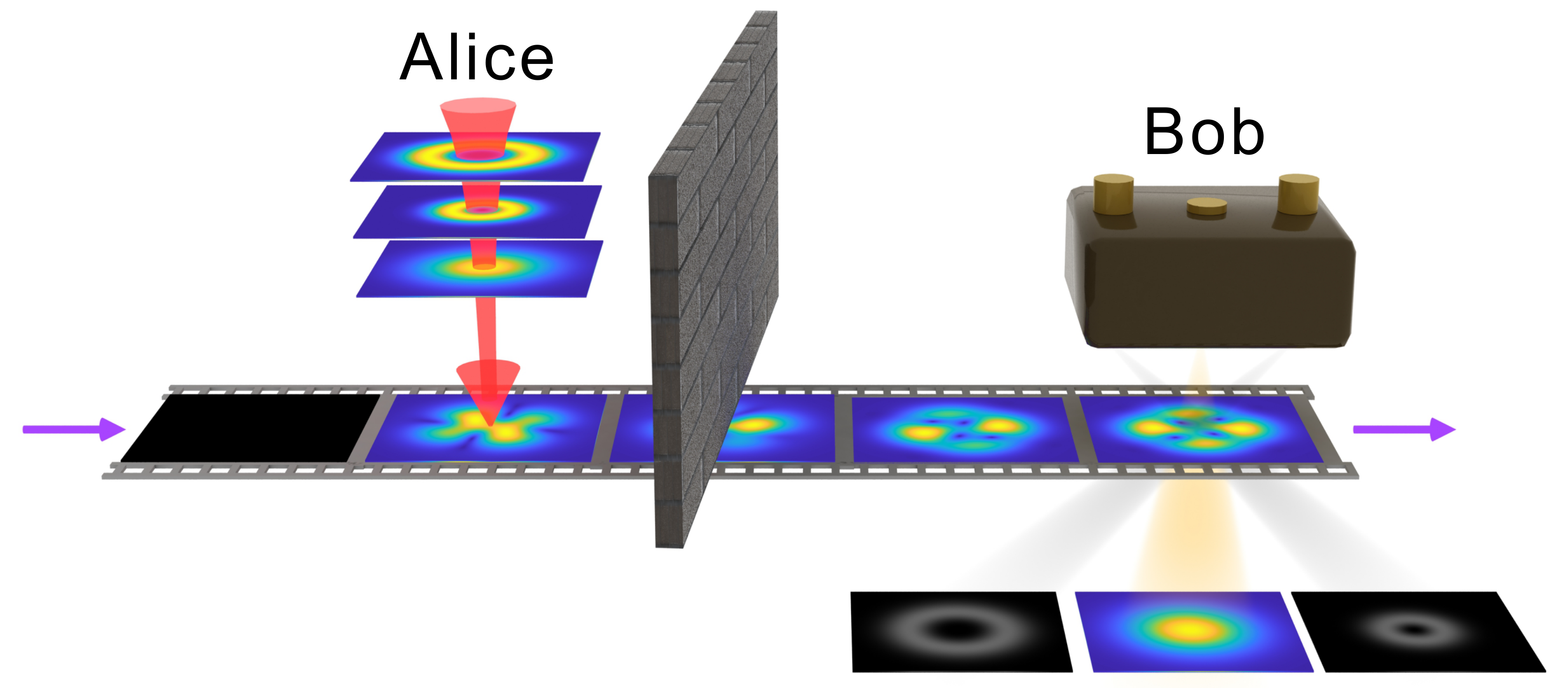}
	\caption{Schematic representation of $n^{(d)}\rightarrow1$ QRACs. The wall represents the restricted channel between Alice and Bob; one qudit and no classical dit can be transmitted per communication. Alice encodes an $n$-dit string into a quantum $d$-level system and sends it to Bob; Bob randomly performs an orthogonal measurement and guesses the corresponding dit from Alice.}
	\label{label_diagram4}
\end{figure}

QRACs are communication tasks involving two separated parties, Alice and Bob. Alice and Bob share a restricted channel; one qudit and no classical dit can be transmitted per communication. As shown in Fig.~\ref{label_diagram4}, Alice is given an $n$-dit string $\bm{x} = x_{0},x_{1},\cdots,x_{n-1}$ chosen uniformly at random and encodes this $n$-dit string into a qudit. Bob is given a number $y$ $\in$ $\mathbb{Z}_n$ ($\mathbb{Z}_n=\{0,1,\cdots,n-1\}$) chosen uniformly at random, and Alice does not know the number $y$. Bob's task is to correctly guess the dit $x_y$ of Alice.


We focus on standard QRACs, which are different from sequential measurement QRACs \cite{Anwer2020}. We care more about the relationship between MUBs and $n^{(d)}{\longrightarrow}1$ QRACs. Therefore, we will introduce the analytical solution for QRACs under the condition of projective measurements \cite{Farkas2019,Aguilar2018,Andris2009}.

First, we construct $n$ fixed orthogonal and complete bases in a $d$-level system: $\left\{ \left| \xi_{i}^{(0)} \right\rangle \right\}$, $\left\{ \left| \xi_{i}^{(1)} \right\rangle \right\}$, $\cdots$, $\left\{ \left| \xi_{i}^{(n-1)} \right\rangle \right\}$ $(i\in \mathbb{Z}_d)$. $\left| \xi_{i}^{(\mu)} \right\rangle$ is the $(i+1)$th state in the $(\mu+1)$th basis. Then we find the best encoding strategy to gain maximum success probability when the bases are fixed.

Alice will encode dits $\bm{x}$ into a qudit $\rho(\bm{x})$. When Bob wishes to obtain the dit $x_y$, Bob will measure the qudit using the $(y+1)$th basis $\left\{ \left| \xi_{i}^{(y)} \right\rangle \right\}$ with success probability $Tr\left( \left| \xi_{x_y}^{(y)} \right\rangle \left\langle \xi_{x_y}^{(y)} \right| \rho(\bm{x}) \right)$; in this case, the success probability of $n^{(d)}\rightarrow1$ QRACs is
\begin{align}
	\bar{P}(\bm{x})=&Tr\left[\sum_{i=0}^{n-1}p_i\left| \xi_{x_i}^{(i)} \right\rangle \left\langle \xi_{x_i}^{(i)} \right| \rho(\bm{x}) \right]\nonumber\\
	=&Tr\left[ \hat{\bar{P}}(\bm{x}) \rho(\bm{x}) \right],\label{label_P_bar_to_hat_bar_P}
\end{align}
where $p_y$ is the probability that Bob chooses dit $x_y$ when Bob wishes to obtain $x_0,x_1,\cdots,x_{n-1}$ not evenly. The operator $\hat{\bar{P}}(\bm{x})=\sum_{i=0}^{n-1}p_i\left| \xi_{x_i}^{(i)} \right\rangle \left\langle \xi_{x_i}^{(i)} \right|$ has $d$ eigenvalues and eigenstates $\{ \lambda_i, \left| \Lambda_i \right\rangle \}$, $\lambda_i\in[0,1)\ (i\in\mathbb{Z}_d)$. Then, we determine the largest eigenvalue $\lambda_m$ and the corresponding eigenstate $\left| \Lambda_m \right\rangle$. Now, the success probability is $\bar{P}(\bm{x})=\sum_{i=0}^{d-1}\lambda_i\left\langle \Lambda_i \right|\rho(\bm{x})\left| \Lambda_i \right\rangle$. When $\rho(\bm{x})=\left| \Lambda_m \right\rangle \left\langle \Lambda_m \right|$, namely, Alice sends a pure state $\left| \Lambda_m \right\rangle$ to Bob, the success probability can reach its maximum point $\bar{P}(\bm{x})=\lambda_m $ \cite{Farkas2019,Aguilar2018,Andris2009}.

Finally, $\lambda_m$ is determined by $\bm{x}$, so the maximum success probability of the $n^{(d)}{\longrightarrow}1$ QRACs when the bases are fixed is
\begin{align}
	P_{nd}^Q=\frac{1}{d^n}\sum_{x_0=0}^{d-1}\sum_{x_1=0}^{d-1}\cdots\sum_{x_{n-1}=0}^{d-1}\lambda_m(\bm{x}).\label{label_PndQ}
\end{align}

Although the optimal measurement bases are MUBs for $2^{(d)}\rightarrow1$ QRACs, the optimal measurement bases need to be found for the maximum success probability of $n^{(d)}\rightarrow1$ QRACs.

\section{\texorpdfstring{Analytical solution for $3^{(\MakeLowercase{d})}{\longrightarrow}1$ QRACs when measurement bases are MUBs}{Analytical solution for 3\^{}(d)->1 QRACs when measurement bases are MUBs}}
For the case of $n=3$, if all bases are mutually unbiased, meaning that $\left|\left\langle \xi_{i}^{(j)} \right. \left| \xi_{k}^{(m)} \right\rangle\right|=1/\sqrt{d}$ when $j\ne m$, then $P_{3d}^Q$ can be simplified. As shown in Fig.~\ref{label_diagram4}, Alice has three dits, and Bob wishes to obtain one of them entirely randomly, i.e., $p_i=1/3\ (i\in\mathbb{Z}_3)$. We construct three fixed MUBs $\left\{ \left | c_i \right \rangle \right\}$, $\left\{ \left | e_i \right \rangle \right\}$, and $\left\{ \left | f_i \right \rangle \right\}$ $(i\in\mathbb{Z}_d)$, where $\left | c_i \right \rangle$ means the $(i+1)$th state in the first basis. Alice encodes $x_0,x_1,x_2$ $\in\mathbb{Z}_d$ into a $d$-level quantum system and sends it to Bob. Here, $\hat{\bar{P}}^M(x_0,x_1,x_2)=\frac{1}{3}\left( \left| c_{x_0} \right\rangle\left\langle c_{x_0} \right| + \left| e_{x_1} \right\rangle\left\langle e_{x_1} \right| + \left| f_{x_2} \right\rangle\left\langle f_{x_2} \right| \right)$. The superscript $M$ represents that measurement bases are MUBs. There are at most three nonzero eigenvalues of $\hat{\bar{P}}^M(x_0,x_1,x_2)$ and corresponding eigenstates must be linear combinations of $ \left| c_{x_0} \right\rangle$, $ \left| e_{x_1} \right\rangle$, and $ \left| f_{x_2} \right\rangle$. It is difficult to reach these eigenstates. However, in the following sections, we can get three states from Eq.~\eqref{label_calculus_psi_theta} and Eq.~\eqref{label_point123_varPhi}. It is unnecessary to know how to get these states here; we only need to know they are all eigenstates of $\hat{\bar{P}}^M(x_0,x_1,x_2)$ with nonzero eigenvalues. These eigenstates $\left| \Lambda_i \right\rangle$ and corresponding eigenvalues $\lambda_i$ of $\hat{\bar{P}}^M(x_0,x_1,x_2)$ are
\begin{align}
	&\begin{cases}
		\lambda_0= \frac{1}{3} \left[ 1+\frac{2}{\sqrt{d}}\cos\frac{\varPhi}{3} \right],\\
		\left| \Lambda_0 \right\rangle = \frac{\left | c_{x_0}  \right \rangle  +  e^{i\left(\frac{\varPhi}{3}-\varphi_{01}\right)} \left | e_{x_1}  \right \rangle  +  e^{i\left(-\frac{\varPhi}{3}-\varphi_{02}\right)} \left | f_{x_2}  \right \rangle}{\sqrt{3+\frac{6}{\sqrt{d}}\cos\frac{\varPhi}{3}}}  ,
	\end{cases}\label{label_eigen_varPhi_origin0}\\
	&\begin{cases}
		\lambda_1= \frac{1}{3} \left[ 1+\frac{2}{\sqrt{d}}\cos\left(\frac{\varPhi}{3}+\frac{2\pi}{3}\right) \right],\\
		\left| \Lambda_1 \right\rangle = \frac{\left | c_{x_0}  \right \rangle  +  e^{i\left(\frac{\varPhi}{3}+\frac{2\pi}{3}-\varphi_{01}\right)} \left | e_{x_1}  \right \rangle  +  e^{i\left(-\frac{\varPhi}{3}-\frac{2\pi}{3}-\varphi_{02}\right)} \left | f_{x_2}  \right \rangle}{\sqrt{3+\frac{6}{\sqrt{d}}\cos\left(\frac{\varPhi}{3}+\frac{2\pi}{3}\right)}},
	\end{cases}\label{label_eigen_varPhi_origin1}\\
	&\begin{cases}
		\lambda_2= \frac{1}{3} \left[ 1+\frac{2}{\sqrt{d}}\cos\left(\frac{\varPhi}{3}-\frac{2\pi}{3}\right) \right],\\
		\left| \Lambda_2 \right\rangle = \frac{\left | c_{x_0}  \right \rangle  +  e^{i\left(\frac{\varPhi}{3}-\frac{2\pi}{3}-\varphi_{01}\right)} \left | e_{x_1}  \right \rangle  +  e^{i\left(-\frac{\varPhi}{3}+\frac{2\pi}{3}-\varphi_{02}\right)} \left | f_{x_2}  \right \rangle}{\sqrt{3+\frac{6}{\sqrt{d}}\cos\left(\frac{\varPhi}{3}-\frac{2\pi}{3}\right)}}.
	\end{cases}\label{label_eigen_varPhi_origin2}
\end{align}
$\varphi_{01}$, $\varphi_{02}$, and $\varphi_{12}$ are the angles of $\left \langle c_{x_0} | e_{x_1} \right \rangle$, $\left \langle c_{x_0} | f_{x_2} \right \rangle$, and $\left \langle e_{x_1} | f_{x_2} \right \rangle$; for example, $e^{i\varphi_{01}}/\sqrt{d} = \left \langle c_{x_0} | e_{x_1} \right \rangle$, and
\begin{align}
    \varphi & =\varphi_{01}-\varphi_{02}+\varphi_{12},\\
	\varPhi & = f\left( \varphi \right).\label{label_varPhi_define}
\end{align}
Here, $f(\varphi)=\varphi-2k\pi$ if $2k\pi-\pi \leqslant\varphi<2k\pi+\pi,\ k\in \mathbb{Z}$, which can be used to transfer the angle $\varphi$ into the range $\left[-\pi,\pi\right)$ by subtracting $2k\pi$. $\varPhi$ is a function of $x_0$, $x_1$, and $x_2$. So $\lambda_0\geqslant\lambda_1,\ \lambda_2$. Therefore Alice sends the best quantum state to Bob and the corresponding success probability is
\begin{align}
	&\bar{P}^M(x_0,\ x_1,\ x_2)= \frac{1}{3} \left[ 1+\frac{2}{\sqrt{d}}\cos\frac{\varPhi}{3} \right],\\
	&\left| \psi_{x_0,\ x_1,\ x_2}^M \right\rangle = \frac{ \left | c_{x_0} \right \rangle + e^{i\left(\frac{\varPhi}{3}-\varphi_{01}\right)} \left | e_{x_1} \right \rangle + e^{i\left(-\frac{\varPhi}{3}-\varphi_{02}\right)} \left | f_{x_2} \right \rangle}{\sqrt{3+\frac{6}{\sqrt{d}}\cos\frac{\varPhi}{3}}}.
\end{align}
Thus, $P_{3d}^Q$ in Eq.~\eqref{label_PndQ} can be simplified to $P_{3d}^{QM}$ when measurement bases are MUBs:
\begin{align}
	P_{3d}^{QM}=\frac{1}{3d^3}\sum_{x_0=0}^{d-1}\sum_{x_1=0}^{d-1}\sum_{x_2=0}^{d-1}\left( 1+\frac{2}{\sqrt{d}}\cos\frac{\varPhi(x_0,\ x_1,\ x_2)}{3} \right). \label{label_P3dQ}
\end{align}
$\varPhi$ cannot be eliminated because $\varPhi$ carries the subset information and $P_{3d}^{QM}$ is related to the choice of subset. The proof of $P_{3d}^{QM}>P_{3d}^C$ (RACs) can be found in Supplemental Material \cite{mySupplementalMaterial}.

\section{\texorpdfstring{Another Way to Get $\bar{P}^M(\MakeLowercase{x}_0,\ \MakeLowercase{x}_1,\ \MakeLowercase{x}_2)$}{Another Way to Get P-\^{}M(x0,x1,x2)}\label{section_calculus}}

There is another way to get the maximum success probability of the $3^{(d)}{\longrightarrow}1$ QRACs when measurement bases are MUBs using the perspective of calculus. We first assume a specific quantum state form and then find the maximum success probability in this form. We prove the rationality of the assumption in the process. Our approach may inspire alternative research tools.

\subsection{Quantum States and Success Probability}

$\left\{ \left | c_i  \right \rangle \right\},\ \left\{ \left | e_i  \right \rangle \right\},$ and $\left\{ \left | f_i  \right \rangle \right\}$ are three fixed MUBs. Alice's data are $\bm{x}=x_0,x_1,x_2$; we assume the quantum state Alice sent to Bob is
\begin{align}
	\left | \psi  \right \rangle=  \frac{1}{N_{3d}}  \left (  \left | c_{x_0}  \right \rangle  + 
	 e^{i\left( \theta_1-\varphi_{01} \right) } \left | e_{x_1}  \right \rangle  +
	 e^{i\left( \theta_2-\varphi_{02} \right) } \left | f_{x_2}  \right \rangle  \right ),\label{label_calculus_psi_theta}
\end{align}
where $\theta_1$ and $\theta_2$ are variables, $N_{3d}$ is a normalization constant, and it is related to $\theta_1$ and $\theta_2$. When Bob wants the first, second, and third data, he performs a measurement in the bases $\left\{ \left | c_i  \right \rangle \right\},\ \left\{ \left | e_i  \right \rangle \right\},$ and $\left\{ \left | f_i  \right \rangle \right\}$; the probabilities that the results are $x_0$, $x_1$, and $x_2$ are $P_{x_0} = {\left | \left \langle c_{x_0}  | \psi  \right \rangle  \right |} ^2$, $P_{x_1} = {\left | \left \langle e_{x_1}  | \psi  \right \rangle  \right |} ^2$, and $P_{x_2} = {\left | \left \langle f_{x_2}  | \psi  \right \rangle  \right |} ^2$.

If Bob chooses $x_0$, $x_1$, and $x_2$ randomly, then we can get the average success probability
\begin{align}
	\bar{P}^M&(x_0,\ x_1,\ x_2)=\frac{P_{x_0}+P_{x_1}+P_{x_2}}{3}\\
	=&\frac{1}{3}  \left\{  2+\left[-3+\frac{6}{d}+\frac{2}{d}\cos\left ( \theta_1-\theta_2 \right )+\frac{2}{d}\cos\left ( \theta_2+\varphi \right )\right.\right.\nonumber\\
	&\left.+\frac{2}{d}\cos\left ( \theta_1-\varphi \right )\right]/\left[3+\frac{2}{\sqrt{d}}\cos \theta_1 + \frac{2}{\sqrt{d}}\cos \theta_2\right.\nonumber\\
	&\left.\left.+ \frac{2}{\sqrt{d}}\cos\left( \theta_1-\theta_2-\varphi \right)\right]  \right\}.
\end{align}

We define a new symbol $q$:
\begin{align}
	\bar{P}^M&(x_0,x_1,x_2)=\frac{2+q}{3},\\
	q=&\left[-3+\frac{6}{d}+\frac{2}{d}\cos\left ( \theta_1-\theta_2 \right )+\frac{2}{d}\cos\left ( \theta_2+\varphi \right )\right.\nonumber\\
	&\left.+\frac{2}{d}\cos\left ( \theta_1-\varphi \right )\right]/\left[3+\frac{2}{\sqrt{d}}\cos \theta_1 + \frac{2}{\sqrt{d}}\cos \theta_2\right.\nonumber\\
	&\left.\left.+ \frac{2}{\sqrt{d}}\cos\left( \theta_1-\theta_2-\varphi \right)\right]  \right\}.
\end{align}

We want to know the maximum value of $\bar{P}^M(x_0,\ x_1,\ x_2)$; in fact, we are calculating the maximum value of $q$. We find, if we replace $\theta_1$ with $-\theta_2$, and replace $\theta_2$ with $-\theta_1$, that $q$ does not change, namely,
\begin{align}
	q\left(\theta_1,\ \theta_2,\ \varphi\right)=q\left(-\theta_2,\ -\theta_1,\ \varphi\right). \label{label_q_fanduichen}
\end{align}

\subsection{\texorpdfstring{Solve $\frac{\partial q}{\partial \theta _1}=0$ and $\frac{\partial q}{\partial \theta _2}=0$}{Solve ∂q/∂θ1=0 and ∂q/∂θ2=0}}

First, we can say the denominator of $q$ is $N_{3d}^2$ and $N_{3d}^2>0$ no matter how $\theta_1$ and $\theta_2$ change because it is the normalization constant for quantum state $\left | \psi  \right \rangle$. Then we conduct a partial derivative operation to $q$:
\begin{align}
	\frac{\partial q}{\partial \theta _1} =&
		-2\sin\frac{2\theta_1-\theta_2-\varphi}{2}\left\{ \cos\frac{\theta_2-\varphi}{2}\left [ \frac{6}{d}+\frac{4}{d^{\frac{3}{2}}}\cos\theta_2 \right ]\right.\nonumber\\
		&\left. + \cos\frac{\theta_2+\varphi}{2}\left [ \frac{6}{\sqrt{d}}-\frac{12}{d^{\frac{3}{2}}}-\frac{4}{d^{\frac{3}{2}}}\cos\left ( \theta_2+\varphi \right )   \right ] \right\}
	/N_{3d}^4.
\end{align}

Using Eq.~\eqref{label_q_fanduichen},
\begin{align}
	\frac{\partial q}{\partial \theta _2} =&
	+
		2\sin\frac{\theta_1-2\theta_2-\varphi}{2}\left\{ \cos\frac{\theta_1+\varphi}{2}\left [ \frac{6}{d}+\frac{4}{d^{\frac{3}{2}}}\cos\theta_1 \right ] \right.\nonumber\\
		&\left. + \cos\frac{\theta_1-\varphi}{2}\left [ \frac{6}{\sqrt{d}}-\frac{12}{d^{\frac{3}{2}}}-\frac{4}{d^{\frac{3}{2}}}\cos\left ( \theta_1-\varphi \right )   \right ] \right\}
	/N_{3d}^4.
\end{align}

The solution of $\frac{\partial q}{\partial \theta_1}=0$ is
	\begin{numcases}{}
		\sin\frac{2\theta_1-\theta_2-\varphi}{2}=0\label{label_partial_1_origin_equation_1}\text{\ \ \ or},\\
		\cos\frac{\theta_2-\varphi}{2}\left [ \frac{6}{d}+\frac{4}{d^{\frac{3}{2}}}\cos\theta_2 \right ] + \cos\frac{\theta_2+\varphi}{2}\left [ \frac{6}{\sqrt{d}}\right.\nonumber\\
		\hspace{0.2in}\left.-\frac{12}{d^{\frac{3}{2}}}-\frac{4}{d^{\frac{3}{2}}}\cos\left ( \theta_2+\varphi \right )   \right ]=0.\label{label_partial_1_origin_equation_2}
	\end{numcases}

The solution of Eq.~\eqref{label_partial_1_origin_equation_1} is
\begin{align}
	2\theta_1-\theta_2-\varphi=2k\pi,\ k\in\mathbb{Z}.
\end{align}

Clearly, Eq.~\eqref{label_partial_1_origin_equation_2} must have at least one solution. There is no need to calculate the solutions because none of them can reach the maximum point in the following steps. We can name any one of the solutions $\gamma_0$; then we can write Eq.~\eqref{label_partial_1_origin_equation_2} as
\begin{align}
	\theta_2=\gamma_0.
\end{align}

\begin{figure}[t]
	\centering
	\includegraphics[width=3.2in]{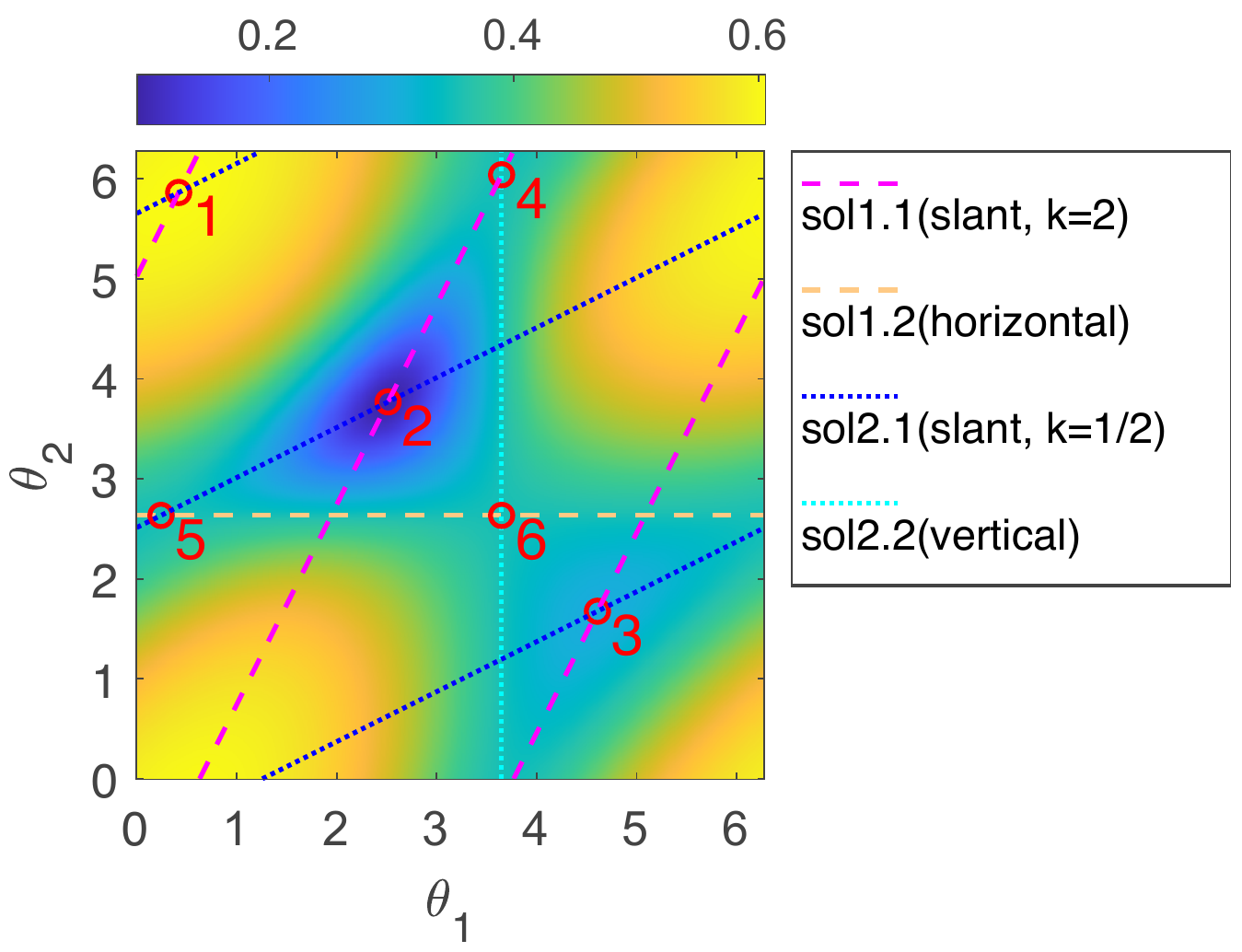}
	\caption{Solutions of $\frac{\partial q}{\partial \theta_1}=0$ and $\frac{\partial q}{\partial \theta_2}=0$. Brightness of each position indicates $q$. Magenta (slant, $k=2$), orange (horizontal), blue (slant, $k=1/2$), and cyan (vertical) lines show sol1.1, sol1.2, sol2.1, and sol2.2, respectively. $q$ of all points on horizontal and vertical lines are the same. Six red points are intersections and the first three of them are intersections between magenta(sol1.1) and blue(sol2.1) lines. Here $d=5$; $x_0,x_1,x_2=0,0,1$; $ \left | e_i \right \rangle = \frac{1}{\sqrt{d}}\sum_{j=0}^{d-1}{\exp(2{\rm{\pi i}}/d)}^{ij} \left| c_i \right\rangle $; and $ \left | f_i \right \rangle = \frac{1}{\sqrt{d}}\sum_{j=0}^{d-1}{\exp(2{\rm{\pi i}}/d)}^{ij+j^2} \left| c_i \right\rangle $. And now $\varPhi\approx1.2566$, $\gamma_0\approx2.6342$.}
	\label{label_q_t1t2_nolegend}
\end{figure}

Then the solution of $\frac{\partial q}{\partial \theta_1}=0$ is
\begin{align}
	\frac{\partial q}{\partial \theta _1}=0\to
	\begin{cases}
		\text{sol1.1:\ \ }\theta_2=2\theta_1-\varphi+2k\pi,\ k\in\mathbb{Z}\text{\ \ \ or},\\
		\text{sol1.2:\ \ }\theta_2=\gamma_0.
	\end{cases}
	\label{label_partial_1}
\end{align}

Using Eq.~\eqref{label_q_fanduichen}, the solution of $\frac{\partial q}{\partial \theta_2}=0$ is
\begin{align}
	\frac{\partial q}{\partial \theta _2}=0\to
	\begin{cases}
		\text{sol2.1:\ \ }\theta_1=2\theta_2+\varphi+2k\pi,\ k\in\mathbb{Z}\text{\ \ \ or},\\
		\text{sol2.2:\ \ }\theta_1=-\gamma_0.
	\end{cases}
\end{align}

The maximum points must be one of the intersections of the above four solutions. The solutions are shown in Fig.~\ref{label_q_t1t2_nolegend}.

\subsection{Find the maximum point}

First, we consider the intersections between sol1.1 and sol2.1:
\begin{align}
	\begin{cases}
		\theta_1=+\frac{\varphi}{3},\\
		\theta_2=-\frac{\varphi}{3},
	\end{cases}
	\begin{cases}
		\theta_1=+\frac{\varphi}{3}+\frac{2}{3}\pi,\\
		\theta_2=-\frac{\varphi}{3}-\frac{2}{3}\pi,
	\end{cases}
	\begin{cases}
		\theta_1=+\frac{\varphi}{3}-\frac{2}{3}\pi,\\
		\theta_2=-\frac{\varphi}{3}+\frac{2}{3}\pi.
	\end{cases}
	\label{label_point123}
\end{align}

Using Eq.~\eqref{label_varPhi_define}, these three intersections are
\begin{align}
	\begin{cases}
		\theta_1=+\frac{\varPhi}{3},\\
		\theta_2=-\frac{\varPhi}{3},
	\end{cases}
	\begin{cases}
		\theta_1=+\frac{\varPhi}{3}+\frac{2}{3}\pi,\\
		\theta_2=-\frac{\varPhi}{3}-\frac{2}{3}\pi,
	\end{cases}
	\begin{cases}
		\theta_1=+\frac{\varPhi}{3}-\frac{2}{3}\pi,\\
		\theta_2=-\frac{\varPhi}{3}+\frac{2}{3}\pi.
	\end{cases}
	\label{label_point123_varPhi}
\end{align}

Actually, these three intersections are shown in Fig.~\ref{label_q_t1t2_nolegend} as points 1, 2, and 3, corresponding to nonzero eigenvalues and eigenstates in Eq.~\eqref{label_eigen_varPhi_origin0}, Eq.~\eqref{label_eigen_varPhi_origin1} and Eq.~\eqref{label_eigen_varPhi_origin2}, showing the rationality of the assumption in Eq.~\eqref{label_calculus_psi_theta}. We define $q_1$, $q_2$, and $q_3$ as values of $q$ at the first, second, and third intersection in Eq.~\eqref{label_point123_varPhi}:
\begin{align}
	q_1&=-1+\frac{2}{\sqrt{d}}\cos\frac{\varPhi}{3},\\
	q_2&=-1+\frac{2}{\sqrt{d}}\cos\left(\frac{\varPhi}{3}+\frac{2}{3}\pi\right),\\
	q_3&=-1+\frac{2}{\sqrt{d}}\cos\left(\frac{\varPhi}{3}-\frac{2}{3}\pi\right).
\end{align}

Here $\varPhi\in \left[-\pi,\pi\right)$, so the maximum point among Eq.~\eqref{label_point123_varPhi} is
\begin{align}
	\begin{cases}
		\theta_1=+\frac{\varPhi}{3},\\
		\theta_2=-\frac{\varPhi}{3}.
	\end{cases}
\end{align}

And the corresponding $q$ is
\begin{align}
	q_{m1}=-1+\frac{2}{\sqrt{d}}\cos\frac{\varPhi}{3}.
\end{align}

Second, we consider all the points on sol1.2, namely, $\theta_2=\gamma_0$, and then use Eq.~\eqref{label_partial_1_origin_equation_2} to get
\begin{align}
	\left.\frac{\partial q}{\partial \theta _1}\right|_{\theta_2=\gamma_0}
	=&-\frac{
		2\sin\frac{2\theta_1-\theta_2-\varphi}{2}}{N_{3d}^4}
	\left\{ \cos\frac{\theta_2-\varphi}{2}\left [ \frac{6}{d}\right.\right.\nonumber\\
	&\left.\left.+\frac{4}{d^{\frac{3}{2}}}\cos\theta_2 \right ]+ \cos\frac{\theta_2+\varphi}{2}\left [ \frac{6}{\sqrt{d}}-\frac{12}{d^{\frac{3}{2}}}\right.\right.\nonumber\\
	&\left.\left.-\frac{4}{d^{\frac{3}{2}}}\cos\left ( \theta_2+\varphi \right )   \right ] \right\}=0.
\end{align}

The last part always equals zero no matter what $\theta_1$ is when $\theta_2=\gamma_0$; at the same time, the last part does not contain $\theta_1$, so $\left.\frac{\partial^n q}{\partial \theta _1^n}\right|_{\theta_2=\gamma_0} =0$ is always right. So when $\theta_2=\gamma_0$, $\theta_1$ has nothing to do with $q$; $q$ is constant; in order to calculate the constant, we can rewrite $q$ as
\begin{align}
	&\resizebox{1\linewidth}{!}{$q=\frac{\left[-3+\frac{6}{d}+\frac{2}{d}\cos\left ( \theta_2+\varphi \right )\right]+\left[ \frac{4}{d}\cos\left ( \frac{\theta_2-\varphi }{2}  \right )\right] \cos\left ( \theta_1-\frac{\theta_2+\varphi }{2}  \right )}
	{\left[ 3 + \frac{2}{\sqrt{d}}\cos \theta_2\right]  + \left[ \frac{4}{\sqrt{d}}\cos\left ( \frac{\theta_2+\varphi }{2}  \right )\right] \cos\left ( \theta_1-\frac{\theta_2+\varphi }{2}  \right )   }.$} \label{label_q_rewrite_theta1}
\end{align}

Now the numerator and denominator both have only one $\theta_1$ item, and we rewrite  Eq.~\eqref{label_partial_1_origin_equation_2} as
\begin{align}
	&\left[ -3+\frac{6}{d}+\frac{2}{d}\cos\left ( \theta_2+\varphi \right )\right]
	\left[ \frac{4}{\sqrt{d}}\cos\left ( \frac{\theta_2+\varphi }{2}  \right )\right] \nonumber\\
	&\hspace{0.26in}=
	\left[ 3 + \frac{2}{\sqrt{d}}\cos \theta_2\right]
	\left[ \frac{4}{d}\cos\left ( \frac{\theta_2-\varphi }{2}  \right )\right] .
	\label{label_theta2_gamma0_transform}
\end{align}

Eq.~\eqref{label_theta2_gamma0_transform} has four parts of Eq.~\eqref{label_q_rewrite_theta1} and they all do not contain $\theta_1$; clearly, $\theta_1$ has nothing to do with $q$, so we can simplify $q$ in the condition of $\theta_2=\gamma_0$ to constant $q_{m2}$, namely,
\begin{align}
	q_{m2}=\frac{-3+\frac{6}{d}+\frac{2}{d}\cos\left ( \gamma_0+\varphi \right )}{3 + \frac{2}{\sqrt{d}}\cos \gamma_0}.
\end{align}

Third, we consider all the points on sol2.2; it is obvious that when $\theta_1=-\gamma_0$ the value of $q$ is also $q_{m2}$ because line $\theta_2=\gamma_0$ and line $\theta_1=-\gamma_0$ has an intersection and $q$ keeps a fixed value on these two lines.

The maximum value is either $q_{m1}$ or $q_{m2}$ and cannot be another value. Then we need to compare $q_{m1}$ with $q_{m2}$; we can transfer $\varphi$ in $q_{m2}$ to $\varPhi$ without changing the value of $q_{m2}$, so let us summarize:
\begin{align}
	q_{m1}&=-1+\frac{2}{\sqrt{d}}\cos\frac{\varPhi}{3},\\
	q_{m2}&=\frac{-3+\frac{6}{d}+\frac{2}{d}\cos\left ( \gamma_0+\varPhi \right )}{3 + \frac{2}{\sqrt{d}}\cos \gamma_0}.
\end{align}

In the case of $\theta_2=\gamma_0$, if $\theta_1\ne-\gamma_0$ and $\theta_1\ne2\gamma_0+\varphi+2k\pi(k\in\mathbb{Z})$, $q=q_{m2}$ but $\frac{\partial q}{\partial \theta _2}\ne0$, so $q_{m2}$ cannot be the maximum value of $q$. So the maximum value of $q$ must be $q_{m1}$.

\subsection{\texorpdfstring{Final Analytical Solution of QRACs when $n=3$}{Final Analytical Solution of QRACs when n=3}}

The maximum point is
\begin{align}
	\begin{cases}
		\theta_1=+\frac{\varPhi}{3},\\
		\theta_2=-\frac{\varPhi}{3}.
	\end{cases}
\end{align}

The maximum value of $q$ is
\begin{align}
	q_{max}=-1+\frac{2}{\sqrt{d}}\cos\frac{\varPhi}{3}.
\end{align}

The corresponding $\bar{P}^M(x_0,\ x_1,\ x_2)$ is
\begin{align}
	\bar{P}^M(x_0,\ x_1,\ x_2)=\frac{1}{3}\left( 1+\frac{2}{\sqrt{d}}\cos\frac{\varPhi}{3} \right).
\end{align}

Here $\varPhi$ carries the information of $x_0$, $x_1$, and $x_2$ and bases $\left\{ \left | c_i  \right \rangle \right\},\ \left\{ \left | e_i  \right \rangle \right\},$ and $\left\{ \left | f_i  \right \rangle \right\}$.

\section{\texorpdfstring{Operational inequivalence of MUBs when dimension is a prime power}{Operational inequivalence of MUBs when dimension is a prime power}}
MUBs play a key role in quantum information processing and have been used in dense coding, teleportation, entanglement swapping, covariant cloning, and state tomography \cite{Durt2010}. It is generally believed that MUBs are maximally incompatible and complementary. However, there are differences even between different subsets of MUBs \cite{Hiesmayr2021}. In the latest research, the OI-MUBs has been discovered \cite{Aguilar2018, Designolle2019}. At present, very little is known about the properties of OI-MUBs. Here, we introduce a pattern from which it is possible to conjecture the OI-MUBs when the dimension $d$ is a prime power, based on $3^{(d)}\rightarrow1$ QRACs.

Now, the set of MUBs in prime power dimension $d$ is said to be complete because there are $d+1$ bases that are all pairwise mutually unbiased  \cite{Wootters1989,Durt2010}. We can choose a three-basis subset of MUBs $\left\{ \left | \xi_{i}^{(\mu)} \right \rangle \right\}\ (\mu\in\mathbb{Z}_{d+1})$ and call them $\left\{ \left | c_i \right \rangle \right\}$, $\left\{ \left | e_i \right \rangle \right\}$, and $\left\{ \left | f_i \right \rangle \right\}$. Then, we can use Eq.~\eqref{label_P3dQ} to obtain the analytical $P_{3d}^{QM}$. After calculating the analytical $P_{3d}^{QM}$ for all subsets when $d\leqslant1000$ ($100$) when dimension $d$ is a prime (prime power) \cite{mySupplementalMaterial}, it is found that the number of different $P_{3d}^{QM}$ ($N$) is no greater than $2$ and conforms to
\begin{equation}
	N = \left\{\begin{array}{ll}
		2, &\text{if}\ d \equiv 1 \ (mod \ 4),\\[0.1cm]
		1, &\mathrm{otherwise,}
	\end{array}\right.
	\ \ (\text{when}\ d \leqslant 1000\ (100)).
	\label{label_NPbar}
\end{equation}

Accordingly, we can find a method to conjecture the OI-MUBs for $3^{(d)}\rightarrow1$ QRACs. When $d \equiv 1 \ (mod \ 4)$, there are two values of $P_{3d}^{QM}$. The larger one is $P_{3d+}^{\text{QM}}$ and the smaller one is $P_{3d-}^{\text{QM}}$, which means that the choice of subset affects $P_{3d}^{QM}$. However, when $d \equiv \text{other} \ (mod \ 4)$, there is only one $P_{3d+}^{QM}$. Our result is obtained using a complete analytical calculation to avoid floating-point errors. we only calculate the $d\leqslant 1000$ $(100)$ case when dimension $d$ is a prime (prime power) because of the lack of computational power, and the computation of the prime power dimension is much more complex than that of the prime dimension. We conjecture Eq.~\eqref{label_NPbar} is also true when $d>1000$ $(100)$. Although the conjecture of OI-MUBs is based on the Galois MUBs, we argue that the conjecture may apply to other completed MUBs.

The maximum success probability of $2^{(d)}{\longrightarrow}1$ QRACs depends only on the dimension \cite{Tavakoli2015}, which means that a case with three MUBs is the simplest case of operational inequivalence. Our results clearly show that the dimension of the Hilbert space plays a central role in the OI-MUBs. If the dimension modulo 4 remains 0, 2, or 3, this means that the Hilbert space has better symmetry. Our results may open alternative avenues for investigating the foundational properties of quantum mechanics.

\section{\texorpdfstring{Galois MUBs are not the optimal measurement bases}{Galois MUBs are not the optimal measurement bases}}

\begin{figure}[t]
	\centering
	\includegraphics[width=3.5in]{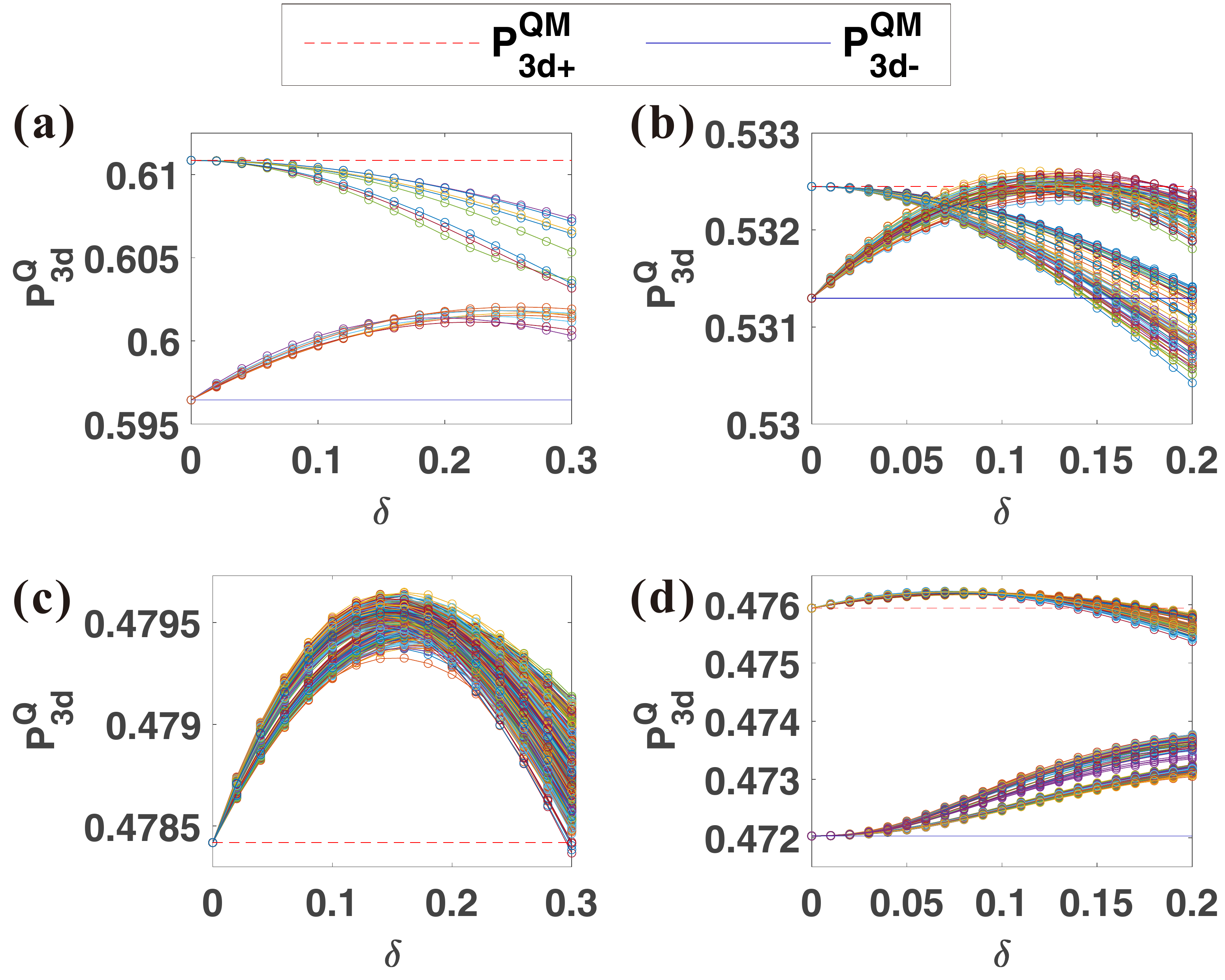}
	\caption{The relationship between MUBs and $3^{(d)}\rightarrow1$ QRACs. (a-d) show $P_{3d}^Q$ for $d=5$, $9$, $16$, and $17$, respectively. $\delta$ is an independent variable in Eq.~\eqref{label_Unew} that represents the degree of deviation from Galois MUBs. The red dashed horizontal line (upper or the only horizontal line) and blue solid horizontal line (lower horizontal line) represent $P_{3d+}^{\text{QM}}$ and $P_{3d-}^{\text{QM}}$. Each curved line (colorful, without legend) represents a choice of a three-basis subset of new bases. The $P_{3d+}^{\text{QM}}$ values for $d=9$, $16$, and $17$ can be surpassed when $d\leqslant20$.}
	\label{fig_noiseP}
\end{figure}

The OI-MUBs already proves that MUBs are not a sufficient condition for maximum success probability of $n^{(d)}\rightarrow1$ QRACs. Then we attempt to find three orthogonal and complete measurement bases that can achieve a greater success probability than any three-basis subset of MUBs when $n=3$. It is very difficult to consider all possible MUBs; here we only focus on the Galois MUBs. Fortunately, such bases are not uncommon in some dimensions. We construct new complete orthogonal quantum bases which we call the perturbations of Galois MUBs \cite{Designolle2019}. First, each quantum state is treated as a complex column vector, and the complete orthogonal basis forms a $d \times d$ matrix $U$ joined by $d$ column vectors. Then, we use perturbation to construct a new basis $U_{\text{new}}$:
\begin{align}
	U_{\text{new}}=SO(U+\delta\mathbb{I}),
	\label{label_Unew}
\end{align}
where $SO$ denotes Schmidt orthogonalization, $\delta$ is an independent variable that represents the degree of deviation from Galois MUBs, and $\mathbb{I}$ is the identity matrix. We transform Galois MUBs to new bases and use each three-basis subset of the new bases to calculate $P_{3d}^Q$ using Eq.~\eqref{label_PndQ}, as shown in Fig.~\ref{fig_noiseP}. Each curved line represents one choice strategy for selecting a three-basis subset from the $d+1$ new bases.

Then, we transform them with different $\delta$ values. For $d=9$ or $17$, we can observe the phenomenon of OI-MUBs in these dimensions, and $P_{3d+}^{QM}$ can be exceeded. Interestingly, in the case of $d=9$, the best measurement bases we obtained come from the bases of $P_{3d-}^{QM}$, whereas in the case of $d=17$ they come from $P_{3d+}^{QM}$. When there is no OI-MUBs, such as $d=16$, we can also exceed $P_{3d+}^{QM}$. However, when $d=5$, we cannot find a surpassing point.

When $n>3$, the OI-MUBs is more complicated \cite{Designolle2019}. We have proved that Galois MUBs are not a necessary condition for maximum success probability of $n^{(d)}\rightarrow1$ QRACs using the method of contradiction. More details can be found in Supplemental Material \cite{mySupplementalMaterial,Mezzadri2007}.

\section{\texorpdfstring{Experiment}{Experiment}}
\begin{figure}[t]
	\centering
	\includegraphics[width=3.3in]{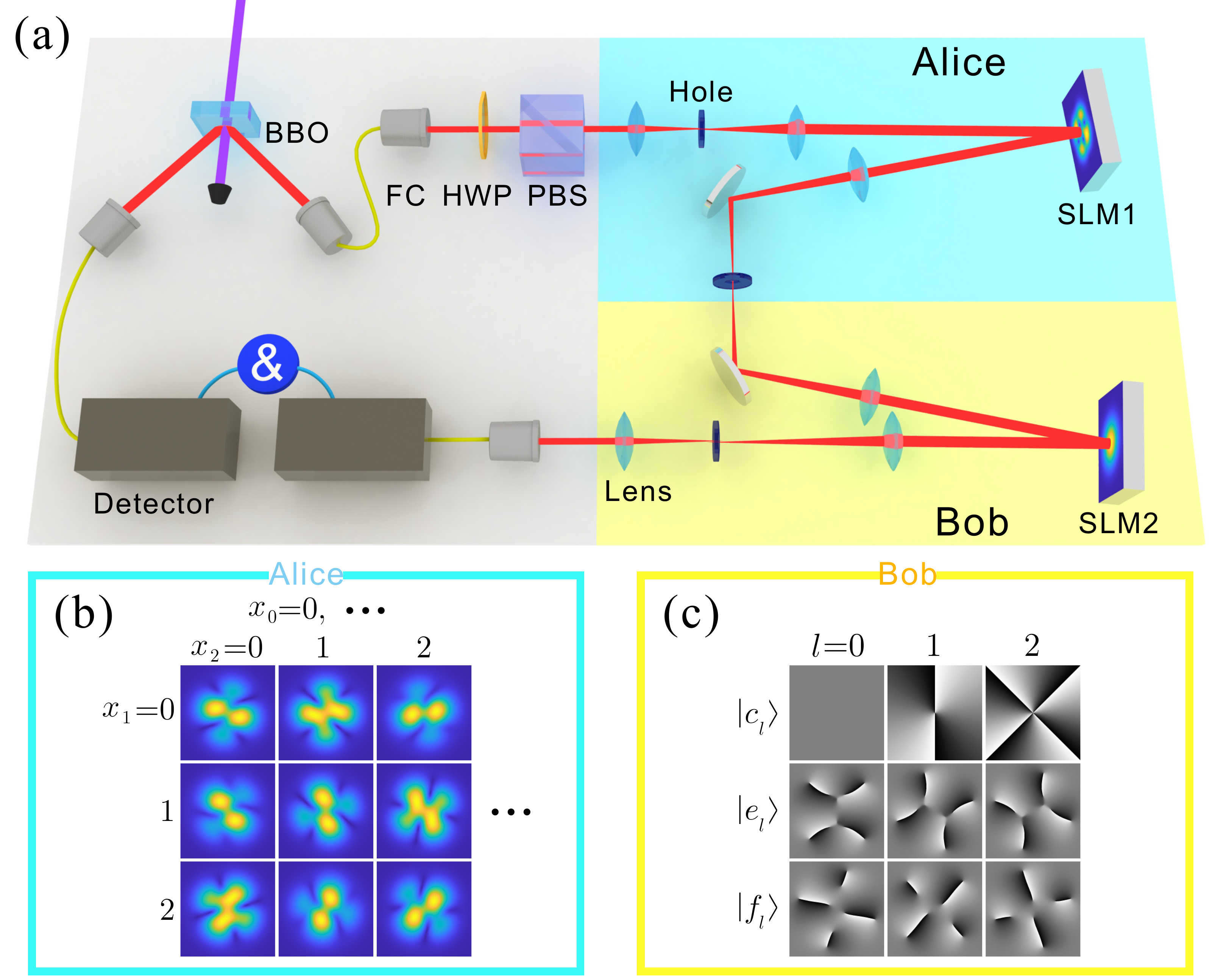}
	\caption{Experimental setup for $n^{(d)}\rightarrow1$ QRACs. (a) We use SLM1 to prepare the quantum state $\left | \psi \right \rangle$ and use SLM2 to measure the state on an orthogonal basis. We show Alice's states and the measurement bases in (b) and (c). Legend: BBO, beta barium borate; FC, fiber coupler; HWP, half-wave plate; PBS, polarizing beam splitter; SLM, spatial light modulator.}
	\label{label_setup}
\end{figure}

In our experiment, as shown in Fig.~\ref{label_setup}, a type-\uppercase\expandafter{\romannumeral2} beta barium borate crystal ($9.0 \times 7.0 \times 1.0$ $mm^{3}$, $\theta = 42.62^{\circ}$) is pumped with a frequency-doubled femtosecond pulse (390-nm, 76-MHz repetition rate) from a mode-locked Ti:sapphire laser to generate single photons. After passing through a 3-nm interference filter, the photon pairs are separately coupled into single-mode fibers. The single-photon state is produced by triggering on one of the two photons.

\begin{figure}[t]
	\centering
	\includegraphics[width=3.2in]{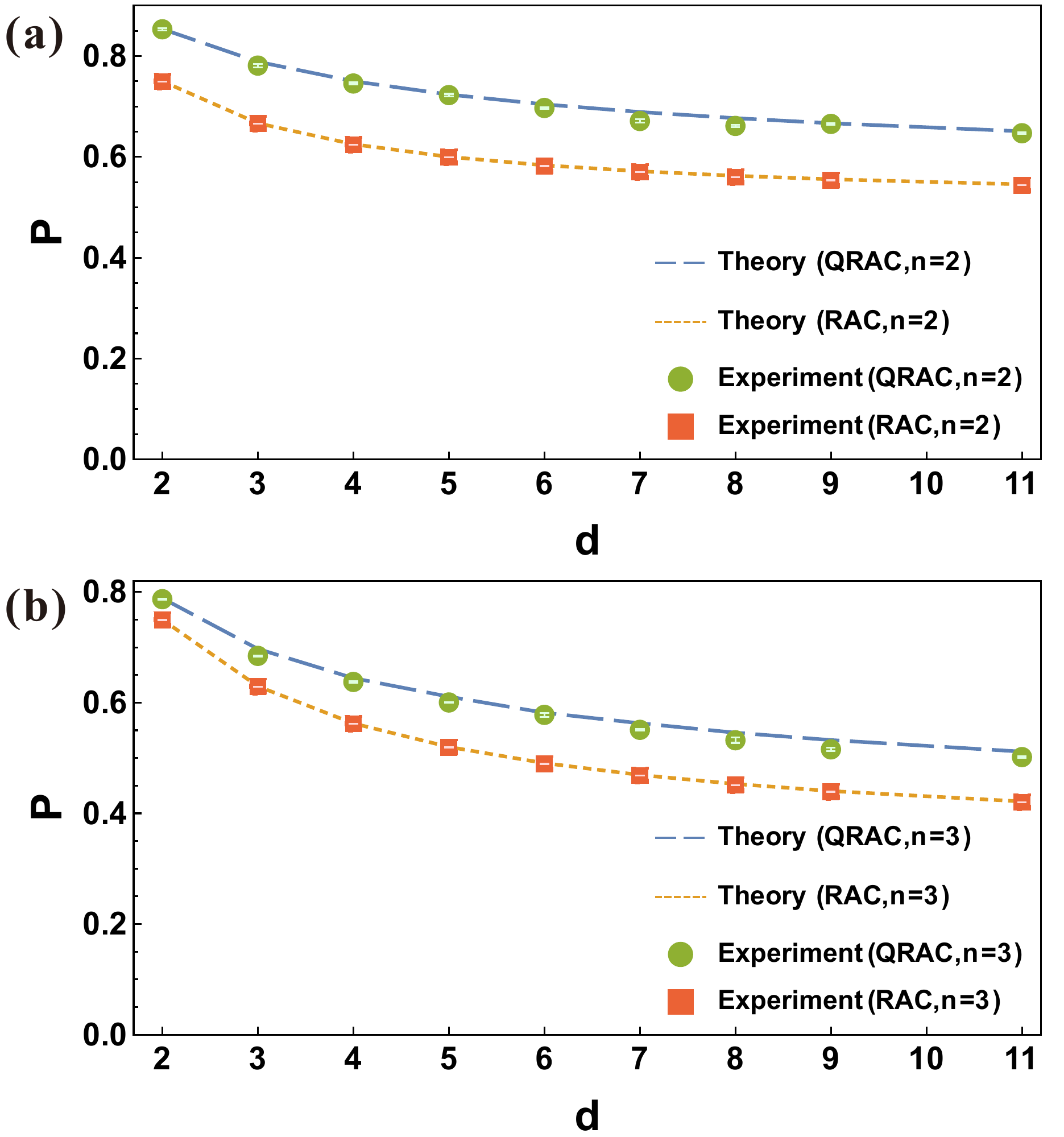}
	\caption{Experimental results for QRACs: (a) $2^{(d)}\rightarrow1$ and (b) $3^{(d)}\rightarrow1$. The standard deviations are shown as white error bars. The largest standard deviation is about $0.0047$, which is too small to be visible.}
	\label{label_resultn_experiment}
\end{figure}

We perform an experiment based on the high-dimensional OAM states of photons (although we use OAM, other high-fidelity high-dimensional quantum systems are also acceptable for QRACs). The prepared states are encoded in the OAM of single photons \cite{Liu2017, Sun2020}. The OAM of single photons can be used to enable high-capacity optical communication \cite{Barreiro2008, Wang2012} and versatile optical tweezers \cite{Padgett2011}. As shown in Fig.~\ref{label_setup}, a signal photon is projected onto the Gaussian state through a single-mode fiber and fiber coupler (FC). The Gaussian diameter of the Gaussian beam after the FC is $2900\mu m$. After spatial filtering and expanding with two lenses, the Gaussian diameter grows to $4450 \mu m$, and the OAM state of the signal photons is controlled by the first spatial light modulator (SLM1) to prepare the desired state $\left | \psi \right \rangle$ for Alice. Many methods have been used to encode high-dimensional states with a single phase-only SLM \cite{Givens1967,32,33}. We calculate the wavefront (including intensity and phase) and adopt a method \cite{33} of modulating the wavefront that is specifically calculated to maximize the state fidelity. Then, the light is passed through a $4f$ system to avoid the Gouy phase-shift effect and reach SLM2. Phase holograms on SLM2 are based on the phase-only wavefront \cite{Frederic2018}. We find that even if SLM1 and SLM2 have a deviation of only $25\mu m$, the fidelity will decrease significantly. Therefore, we generate phase holograms larger than SLMs and move them on the computer instead of moving a real SLM device. To align two SLMs, we put a CCD near the final FC and draw $l$-vortex and $g$-grating holograms ($\varphi(x,y)=l\times angle(x,y)+2\pi x/g$) on two SLMs, where $g$ is the same as in the previous method \cite{33}. When $l=5$, we can see a halo on the CCD. Then if we successfully align two SLMs, there will be a bright spot in the center of the halo, and the moving hologram will cause the bright spot to move. Finally, after passing through another $4f$ system, the light is collected by a single-mode fiber with an adjustable FC and a single-photon detector. The coincidence counting rate collected by the avalanche photodiodes is approximately $6000$ per second if holograms on two SLMs are grating holograms. When the holograms are preparation and measurement holograms, the coincidence counting rate changes when $d$ and Alice's data $\bm{x}$ change; when $d=5$ it is about $1000$ per second. The time for each measurement process is 5 s. We repeat each experiment five times to obtain the standard deviation, which is almost the same as the standard deviation calculated by the Monte Carlo algorithm.

\begin{figure}[t]
	\centering
	\includegraphics[width=3.2in]{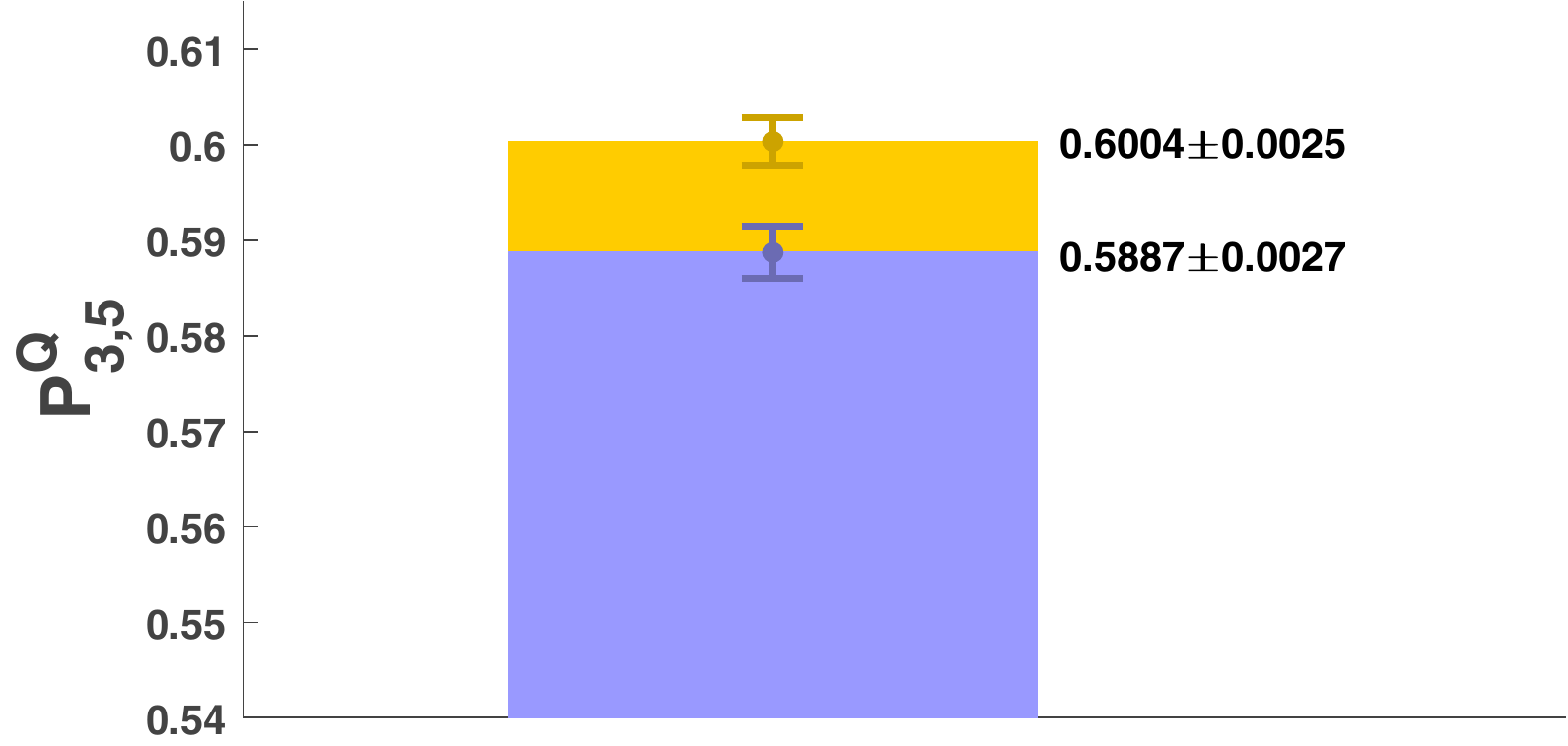}
	\caption{Experimental results for the OI-MUBs when $d=5$. The orange (upper) and blue (lower) rectangles represent the experimental results in the $P_{3d+}^{\text{QM}}$ and $P_{3d-}^{\text{QM}}$ situations, respectively. The error bars of $P_{3d+}^{\text{QM}}$ and $P_{3d-}^{\text{QM}}$ are displayed in the middle of the top edge of rectangles.}
	\label{label_PmaxPminExperiment}
\end{figure}

We first experimentally investigate high-level QRACs up to dimension 11 based on the high-fidelity high-dimensional OAM states. As shown in Fig.~\ref{label_resultn_experiment}, we realize $2^{(d)}\rightarrow1$ and $3^{(d)}\rightarrow1$ QRACs with dimensions of up to 11 (except 10). We need high-dimensional bases, so we use both the azimuthal index $l$ and the radial index $p$ of the Laguerre-Gaussian mode. Let us take $d=11$ as an example. We use $\left| l=0,\pm 2,\pm 4,\pm 6;p=0 \right\rangle$ and $\left| l=\pm 1,\pm 3;p=2 \right\rangle$ as the basic basis $\left\{ \left| \xi_{i}^{(0)} \right\rangle \right\}$; the visibility of $\left\{ \left| \xi_{i}^{(0)} \right\rangle \right\}$ is $99.25\%$ (namely, 132:1) \cite{mySupplementalMaterial}. Then we construct Galois MUBs \cite{Wootters1989,Durt2010}. Fig.~\ref{label_resultn_experiment} (a) represents $2^{(d)}\rightarrow1$ QRACs; the $n=2$ case does not have OI-MUBs so we do not need to consider the choice of subset of MUBs. Fig.~\ref{label_resultn_experiment} (b) represents $3^{(d)}\rightarrow1$ QRACs. For all kinds of dimensions except $d=9$, considering the OI-MUBs, we choose the best subset of Galois MUBs which can reach the success probability of $P_{3d+}^{QM}$. When $d=9$, we choose the bases of the maximum point in Fig.~\ref{fig_noiseP} (b). These bases are shown with three decimal places precision in Supplemental Material \cite{mySupplementalMaterial}. The standard deviations are shown as white error bars. The largest standard deviation in our experiment is about $0.0047$, too small to be visible in Fig.~\ref{label_resultn_experiment}. We can see the experimental results are in good agreement with the theory and they are shown in Supplemental Material \cite{mySupplementalMaterial}.

Finally, we experimentally verify the OI-MUBs. When $d=5$, $P_{3d+}^{QM} \approx 0.6109$ and $P_{3d-}^{QM} \approx 0.5964$. We measure the $P_{3d+}^{QM}$ and $P_{3d-}^{QM}$ individually (as shown in Fig.~\ref{label_PmaxPminExperiment}). Excitingly, our results can be discriminated even when the standard deviation is considered. For the $P_{3d+}^{QM}$, the success probability we obtain is $0.6004 \pm 0.0025$, which is larger than both the theoretical value ($P_{3d-}^{QM}$) and the experimental value $0.5887 \pm 0.0027$ in the $P_{3d-}^{QM}$ situation. The gap is about $4.33\ \sigma$. Although the interference in real experiments leads to the increase of standard deviation, we can see that it is feasible to realize the OI-MUBs. The OI-MUBs is a remarkable phenomenon that cannot be ignored in a real experiment.

\section{\texorpdfstring{Conclusion}{Conclusion}}
In this paper, we focus on $n^{(d)}\rightarrow1$ QRACs. The general method for the maximum success probability of $n^{(d)}\rightarrow1$ QRACs is given. When measurement bases are MUBs, we obtain an analytical solution for the maximum success probability of $3^{(d)}\rightarrow1$ QRACs. Based on this analytical solution, we obtain some interesting results. First, we present a pattern from which it is possible to conjecture the OI-MUBs when $d$ is a prime power. We provide numerical proof of such operational inequivalence up to dimension 1000 (100) when $d$ is a prime (prime power). Then, we find that Galois MUBs are not the optimal measurement bases in contrast to the traditional concept. Considering the three-basis subset of Galois MUBs cannot cover all the MUBs triplets, so it is still an open question to investigate whether MUBs are the optimal measurement bases. Finally, we experimentally achieve $2^{(d)}\rightarrow1$ and $3^{(d)}\rightarrow1$ QRACs, with a dimension of up to 11. In particular, because of our high fidelity, we experimentally confirm the OI-MUBs when $d=5$. The current framework not only brings light to the study of QRACs and MUBs properties but also can be extended to multiqudit and sequential measurement cases. Our results open alternative avenues for investigating quantum network coding.

\begin{acknowledgments}
We are grateful to William Wootters, Edgar A. Aguilar, Oliver Reardon-Smith, and M\'at\'e Farkas for their discussions and help regarding this paper. This work was supported by the Innovation Program for Quantum Science and Technology (Grant No. 2021ZD0301200), the National Natural Science Foundation of China (Grants No. 62005263, No. 62105086, and No. 11821404), and the China Postdoctoral Science Foundation (Grant No. 2020M671862).
\end{acknowledgments}


\nocite{*}

\bibliographystyle{apsrev4-2}
\bibliography{apssamp}

\end{document}